**Social Contagion Theory:**
**Examining Dynamic Social Networks and Human Behavior**

Nicholas A. Christakis[1,2*], James H. Fowler[3,4]

[1]*Department of Medicine and Department of Health Care Policy, Harvard Medical School, Boston, MA 02115*

[2]*Department of Sociology, Faculty of Arts and Sciences, Harvard University, Cambridge, MA 02138, USA*

[3]*Division of Medical Genetics, University of California, San Diego, La Jolla, CA 92093, USA*

[4]*Department of Political Science, University of California, San Diego, La Jolla, CA 92093, USA*

* To whom correspondence should be addressed, email: christakis@hcp.med.harvard.edu.

**Abstract**:

Here, we review the research we have done on social contagion. We describe the methods we have employed (and the assumptions they have entailed) in order to examine several datasets with complementary strengths and weaknesses, including the Framingham Heart Study, the National Longitudinal Study of Adolescent Health, and other observational and experimental datasets that we and others have collected. We describe the regularities that led us to propose that human social networks may exhibit a "three degrees of influence" property, and we review statistical approaches we have used to characterize inter-personal influence with respect to phenomena as diverse as obesity, smoking, cooperation, and happiness. We do not claim that this work is the final word, but we do believe that it provides some novel, informative, and stimulating evidence regarding social contagion in longitudinally followed networks. Along with other scholars, we are working to develop new methods for identifying causal effects using social network data, and we believe that this area is ripe for statistical development as current methods have known and often unavoidable limitations.

Acknowledgements: We thank Weihua An, Felix Elwert, James O'Malley, JP Onnela, Mark Pachucki, and Alan Zaslavsky for helpful comments on the manuscript, and we thank colleagues with whom we have written papers on social networks cited here.

In 2002, we became aware of the existence of a source of raw data that had not previously been used for research purposes. While limited in certain ways, these data offered important strengths and opportunities for the study of social networks. As described below, we were able to exploit previously unused paper records held by the Framingham Heart Study (FHS), a longstanding epidemiological cohort study, to reconstruct social network ties among 12,067 individuals over 32 years. In particular, a very uncommon feature of these data was that the network ties themselves were longitudinally observed, as were numerous attributes of the individuals within the network. We called the resulting dataset the "FHS-Net."

In 2007, we began to publish papers using this dataset – and also other datasets, including the National Longitudinal Study of Adolescent Health (AddHealth, a public-use dataset with social network information on 90,000 children in 114 schools) [1], online social network data that we extracted on both a small [2] and large [3] scale, *de novo* data that we have collected regarding populations as diverse as American college students and Hadza hunter-gatherers [4, 5], and experimental data in which interaction networks or influence paths were artificially created [3, 6, 7] – in order to examine various network phenomena. These datasets have complementary strengths and weaknesses, as do the various analytic approaches we have employed.

There are two broad classes of investigations of networks that we have undertaken: studies of network topology (and its determinants), and studies of the spread of phenomena across network ties. While we have done work on the former [5, 7-13], here we will focus primarily on the latter, discussing analyses of the flow of behaviors, affective states, or germs. Our work on social networks and human behavior thus covers several domains and relies on diverse data and approaches. It builds on prior research on "peer effects" and inter-personal influence by examining data in which individuals are embedded in networks much larger than two people. Here, we summarize this work and describe critiques, extensions, and confirmations of our findings by other scientists.

Using similar modeling approaches and exploiting data from many sources, we have examined the "spread" of obesity [14, 15], smoking [16], alcohol consumption [17], health screening [18], happiness [19], loneliness [20], depression [21], sleep [22], drug use [22], divorce [23], food consumption [24], cooperative behavior [6], influenza [4], sexuality and sexual orientation [25], and tastes in music, books, and movies [26]. We have also conducted experiments regarding the spread within networks of altruism [6, 7] and of political mobilization [3]; in such experiments, causal inference with respect to network effects is more robust (though experiments have limitations of their own). We have previously summarized this work, and also the work of numerous other scholars who have investigated social networks and interpersonal influence, in our book, *Connected*, published in 2009 [27], and in a 2008 review article focused on health [28].

In our work, we have used the best currently available methods. Network statistics is a fast-growing field (for useful reviews of the topic, see: [29-36]), and it is clear that perfect methods, free of any limitations or assumptions, do not exist for every sort of question one might want to ask with observational (or even experimental) data. Basic issues in coping with missing data (missing nodes, ties, covariates, waves), sampling (design effects and incomplete network ascertainment), computation of standard errors, and even of the causal interpretation of model parameters, for example, are still being addressed.

But rather than foreswear observations regarding social network phenomena, we have chosen, in our papers, to analyze available data, and we attempt to characterize known limitations and assumptions in available methods. And, of course, as scientists identify limitations in current methods, many will, we hope, also take the next step to innovate and propose alternatives, since all statistical methods have limitations and they frequently rely on untestable or awkward assumptions. We hope our own work has played a part in stimulating interest in developing statistical methods for network data; we are interested to deploy new and better methods, and we are attempting to contribute to progress in this area, as described below. Hence, we invite suggestions regarding how to analyze such data if current approaches have limitations that some find overwhelming.

This paper proceeds as follows. First, in Section (1) we describe a key dataset that we assembled and first analyzed, the so-called FHS-Net. While we describe the FHS-Net in detail, we note that we and others have replicated our findings using other datasets and methods, as discussed below, including by using experiments. In Section (2) we describe basic analyses involving permutation tests that show clustering of various traits within various observed social networks. Section (3) addresses a set of concerns regarding the nature of potential biases introduced to estimates of clustering by the limited nature of social ties available in the FHS-Net. Section (4) describes the longitudinal regression models we deployed to analyze peer effects within the network, at the dyadic level. We attempt to provide a comprehensive review of the assumptions and biases present in such models. And we summarize model output as applied to more than one dataset. In Section (5), we describe a novel identification strategy we proposed in 2007 involving the exploitation of the directionality of some social ties. We also describe extensions and limitations since characterized by other scientists. Section (6) describes how geographic location information might be used to help address certain types of confounding with observational network data. Section (7) describes how the FHS-Net data has been publicly available since 2009, and where other data regarding longitudinally evolving networks might also be obtained. Section (8) concludes and also summarizes much work that has been done in recent years by other scholars documenting spreading processes in networks.

(1) The FHS-Net Data and Its Pertinent Features

We start by describing a key (but not the only) dataset that motivated our work. The Framingham Heart Study was initiated in 1948 when 5,209 people in Framingham, Massachusetts, were enrolled into the "Original Cohort" [37]. In 1971, the "Offspring Cohort," composed of most of the children of the Original Cohort, and their spouses, was enrolled [38]. This cohort of 5,124 people has had almost no loss to follow-up other than because of death (only 10 cases in the Offspring Cohort dropped out and were un-contactable by the study managers, and there was a similarly low loss to follow-up in the other cohorts). In 2002, enrolment of the so-called "Third Generation Cohort" began, consisting of 4,095 of the children of the Offspring Cohort. The Framingham Heart Study also involves certain other smaller cohorts. Participants in all the cohorts come to a central facility for detailed physical examinations and data collection every 2-4 years.

For many decades, the FHS has maintained handwritten tracking sheets that administrative personnel have used to identify people close to participants for the purpose of facilitating follow-up. These documents contain valuable, previously unused social network information because they systematically (and, in some cases, comprehensively) identify relatives, friends, and co-workers of study participants. To create the network dataset, we computerized information about the Offspring Cohort from these archives.

In the field of social network analysis, procedures for identifying social ties between individuals are known as "name generators" [39, 40]. The ascertainment of social ties in the FHS was both wide and systematic. The FHS recorded complete information about all first-order relatives (parents, spouses, siblings, children), whether alive or dead, and also at least one "close friend" (the set-up and question asked were "please tell us the name of a close friend, to whom you are not related" with whom "you are close enough that they would know where you are if we can't find you"). This information was collected at each of seven exams between 1971 and 2003. Detailed home address information was also captured at each time point, and we computerized and geocoded it. Information about place of employment at each wave allowed us to identify ties to co-workers within the network (by seeing whether two people worked at the same place at the same time). As noted below, all of a person's contacts of the foregoing types were recorded, whether those contacts were also themselves participants in one of the FHS cohorts, and we computerized all this information.

Over the course of follow-up, the participants spread out across the USA, but they nevertheless continued to participate in the FHS. As a person's family changed due to birth, death, marriage, or divorce, and as their contacts changed due to residential moves, new places of employment, or new friendships, this information was captured. For any given "ego" (the person of interest) in the data, a particular "alter" (a person who has a relationship with the ego) may usually be placed in one mutually exclusive category: spouse, sibling, parent, friend, and so on, though, depending on the analysis, we can also allow multiple categories (for example, a co-worker or neighbor might be a friend or sibling). Further details about our data development process are available in our published work.

We used the Offspring Cohort as the source of 5,124 egos to study. Any persons to whom these subjects were linked in any sort of relationship – in any of the FHS cohorts, including the Offspring Cohort itself – can serve as alters. In total, there were 12,067 egos and alters across all cohorts of the FHS who were connected at some point during 1971 to 2003.

We observed ties to individuals both inside and outside the sample. For example, an ego might be connected to two siblings, one of whom was also a participant in the FHS and one of whom was not. For those who were also participants, we could observe their attributes (for example, their health status) longitudinally. Overall, as of 2009 and wave 7 of data collection, there were 53,228 observed familial and social ties to the 5,124 subjects observed at any time from 1971 to 2009, yielding an average of 10.4 ties per subject within the network (*not* including ties to residential neighbors). Fully 83% of subjects' spouses were also in the FHS and 87% of subjects with siblings had at least one sibling in the FHS. We also know the identity of spouses, siblings, and other contacts who are outside our sample; and, though they are not in the FHS, we have basic information about them (e.g., their residential location and vital status).

Importantly, 45% of the 5,124 subjects were connected via friendship to another person in the FHS at some point, which allowed us to observe outcomes for both the naming friend and the named friend. In total, there were 3,542 such friendships for an average of 0.7 friendship ties per subject. For 39% of the subjects, at least one coworker was captured in the network at some point. For 10% of the subjects, an immediate (non-relative) residential neighbor was also present (more expansive definitions, such as living within 100 meters, resulted in many more subjects having identifiable "neighbors").

Our published papers and supplements contain detailed analyses of the possible biases in terms of who among an ego's alters are in and out of the network sample. In general, the pattern is one of limited difference. Egos who name social contacts who are also participants in the FHS are not significantly different from those whose contacts are not in the FHS with respect to their weight [14], smoking behavior [16], alcohol consumption [17], happiness [19], loneliness [20], or depression [21].

The types of alters that we identified for each ego, the number we identified, and the number we were able to also include in our actual sample, are generally not far from data gathered on unrestricted, national samples. For instance, work by others using the General Social Survey identifies the size of people's 'core discussion group' as being about 4-6 people, including one's spouse, siblings, friends, and so on [41, 42]. In our own work with a representative sample of 2,900 Americans collected in conjunction with a Gallup Organization, we find that, on average, in response to the commonly used name generators ("Who do you spend free time with" and "Who do you discuss important issues with"), Americans identify an average of $4.4 \pm 1.8$ alters. And, the average respondent lists 2.2 friends, 0.76 spouses, 0.28 siblings, 0.44 coworkers, and 0.30 neighbors who meet these name-generator criteria [43]. Finally, although the FHS is almost exclusively white and tends to have somewhat more elevated education and income than a representative group of Americans, it appears that subjects' health-related attributes are similar to broader populations of Americans.

The FHS-Net underwent ongoing development over the course of our work, and we are still upgrading it. For instance, co-worker ties were not available in 2007, but were by 2008; an 8th wave of data collection has recently become available; and we have extended the number of individuals about whom we have geocoding and network data substantially.

## (2) Basic Analyses and Findings: Clustering

One of the first types of computations we performed with most of the network phenomena we have studied involved assessing whether there was more "clustering" of a trait of interest (that is, co-occurrence of the trait in connected individuals) than might be expected due to chance alone.[1] To do this, we implemented the following permutation test: we compared the observed clustering in the network to the clustering in thousands of randomly generated networks in which we preserved the network topology and the overall prevalence of the trait of interest, but in which we randomly shuffled the assignment of the trait value to each node in the network [44] [45].

That is, for any given time interval (e.g., for a survey wave), the network topology is taken as static. Then, nodes are randomly assigned to have the trait of interest, subject to the constraint that the prevalence of the trait is fixed. This is done repeatedly. The statistic that is then calculated, for each geodesic distance, is the percentage increase in the probability (i.e., a risk ratio) that an ego has the trait of interest given that an alter also has the trait, compared to the probability that an ego has the trait of interest given that the alter does not. If clustering is occurring, then the probability that an alter has a trait of interest (e.g., obesity) given that an ego has the trait should be higher in the observed network than in the random networks.[2] This procedure also allows us to generate the range of values that might occur due to chance (with 95% probability), and we show these ranges as confidence intervals around the observed value (specifically, we show the distribution of the observed value *minus* the permuted values). This permutation test thus provides a way to test the null hypothesis that the observed value minus the permuted value is equal to zero. If the range crosses zero, it means that the observed value falls between the 2.5[th] and 97.5[th] percentile of the permuted values and we cannot reject the possibility that the observed value could have arisen due to chance.[3]

---

[1] This type of "clustering" is not the same as another frequently described type of clustering in network science, namely, the "clustering coefficient," which captures the degree to which two people tend to share the same social connections.

[2] It is worth noting that, as executed, the null distribution is a completely random distribution of the pertinent traits on the network. This allows us to reject the most simple of models. But it does not demonstrate that the data are more clustered than predicted based on, for example, homophily on age or on other attributes that one might want to hold in place while examining phenomena of interest. Moreover, there could be still more complex assumptions,

Thus, we can measure how far, in terms of geodesic distance (i.e., the number of steps taken through the network), the correlation in traits between ego and alters reaches before it could plausibly be explained as a chance occurrence. In many empirical cases, we found that this relationship extended up to three degrees of separation. In other words, on average, there is a statistically significant and substantively meaningful relationship between, say, the body mass index (BMI, which is weight divided by height squared, in units of $kg/m^2$) of an ego and the BMI of his or her friends (geodesic distance 1), friends' friends (geodesic distance 2), and even friends' friends' friends (geodesic distance 3).

At least one author has raised the concern that incomplete ascertainment of the network could be driving these results, since we might not know that friends' friends' friends were, in actuality, directly connected to an ego [47]. But other data sets with more complete ascertainment of ties than the FHS-Net still often show clustering to three degrees. Figure 1 demonstrates significant clustering up to three degrees for various outcomes in the FHS-Net and also in other data sets, such as AddHealth, Facebook, and even laboratory experiments [6]. Many of these data sets have virtually complete network ascertainment, capturing all relevant ties. This suggests that censoring of out-degree is not the sole source of the conclusions drawn from analyses of the FHS-Net. Moreover, we find similar effect sizes in terms of obesity in both the FHS-Net and AddHealth [15]. Finally, as discussed below, work by other groups with diverse datasets has confirmed our findings; and, in any case, as also discussed below, incomplete sampling would not perforce inflate estimates of geodesic distance.

Permutation tests like this, to test whether a set of observations can result from a chance process, are widely used when the underlying distribution is unknown. There is a substantial literature on this technique, starting with R. A. Fisher [48]; the technique has been applied to networks by other scholars, [44] and it is fairly widely used in network science research. At least one commentator has suggested that this approach is generically limited [49]. We would certainly welcome a closed-form test with well-understood asymptotic properties, but the network dependencies make such a test difficult to describe analytically, and we invite suggestions regarding alternatives.

Now, as explicitly noted in all our papers, there are three explanations (other than chance) for clustering of individuals with the same traits within a social network: (1) subjects might choose to associate with others exhibiting similar attributes (*homophily*) [50]; (2) subjects and their contacts might jointly experience unobserved contemporaneous exposures that cause their

---

such as an assumption that persons with a particular trait have higher degree. But the possible specification of such null models is very broad. And developing such tests is not a trivial exercise. In the supplements to some of our papers, we do evaluate whether clustering in the networks is occurring above and beyond homophily on certain attributes, such as education, by using adjusted values (we take the residual value from a regression that includes the attribute and treat this as the outcome of interest).

[3] An alternative way to present the same information is to show the permuted range around zero, and then test the null hypothesis that the observed value falls inside the permuted range. See [46] for some recent, additional exposition of such issues.

attributes to covary (omitted variables or *confounding* due to shared context); and (3) subjects might be influenced by their contacts (*induction*). But this observation is nothing new.[4]

All observational studies seeking to estimate causal processes must cope with the fact that correlations may result from selection effects or spurious associations instead of a true causal relationship. Correlation is – of course – not causation. But this does not mean that any observational evidence is uninformative. The challenge is to disentangle these effects, to the extent possible, and to specify the assumptions needed before correlative evidence can be taken as suggestive of causation. Distinguishing inter-personal induction from homophily is easier when (subject to certain statistical assumptions) longitudinal information both about people's ties and about their attributes (*i.e.*, obesity, smoking) is available [51, 52], or when certain other techniques (such as the directional test described below) are used. Of course, actual experimental data helps a lot here too, as in [3, 6, 7, 53, 54].

To be clear, what the observed values and confidence intervals from the permutation test described above establish is this: if we do not know anything about a subject in a network except one fact – that his friend's friend's friend has some attribute – then we can do better than chance at predicting whether or not the subject has the same value of the attribute. Of course, it is unclear whether this simple, uncontrolled association results from influence (spread), homophily, contextual factors, or some combination of these, and this is why further analytic approaches are needed.

To illustrate the baseline clustering that exists in the populations we study, we usually present at least one image of the network that shows each individual's characteristics (behavior, gender, and so on) and the social relationships they have. In most cases with large datasets like ours, it is difficult to show the full network because it would be too intricate, so we usually show only a part of the network. Two illustrative examples are in Figure 2. For example, we either carefully select which kinds of social relationships to include (as we did in our image of obesity) or we sample subjects (as we did in our image of happiness), and we show a fully connected "component" (every subject has a relationship with at least one other subject in the group). We choose the largest component, which allows inspection of individual relationships while still conveying the complexity of the overall data. We have used the same techniques to choose subjects to include in movies of the network that show how it evolves and changes over time, prepared with SONIA (examples of such videos are available at our websites) [55].

We sometimes employ a technique we call "geodesic smoothing" to make it easier to see large-scale structure in the network. In this technique, we color each node according to the average value of a characteristic (e.g., happiness) for a person and all of the person's direct social contacts. This process is analogous to smoothing algorithms like LOESS that are used to show trends in representations of noisy data. Geodesic smoothing tends to make it easier to visualize clusters with distinct characteristics, but, since it generates additional correlation between the nodes in the network, we never use these values in our simulations or our statistical tests. They are generated only for the purpose of visualization. In all cases, these techniques are explicitly described in our papers. Similar techniques are described elsewhere [36, 56].

---

[4] In fact, these issues were identified in the 19[th] century, when the study of the widowhood effect was first engaged (the widowhood effect is a simple, dyadic, inter-personal health effect, and it is quite likely the earliest example of social network health effects to receive scholarly attention, as discussed in *Connected*). Moreover, it is worth noting that all three of these phenomena are typically present in most social processes. To be clear, it is not necessary for scholars to set up a false dichotomy – namely, that there is *either* homophily *or* influence in some process. Both are typically always present (though there are obvious exceptions, e.g., that homophily on race in friendships is not due to influence whereby one person's race causes a change in the other's). And, different analysts might be focused on different phenomena, depending on their interests. Some might be interested in exogenous factors that cause people to form ties or share an attribute; others might be interested in how sharing an attribute causes people to form a connection; and still others will be interested in inter-personal influence. Depending on the analyst's interest, the other phenomena will be nuisances that must be dealt with in estimation.

Some people unfamiliar with network visualizations have formed the impression that they are entirely arbitrary. But they are not. The pictures are visual heuristics, and it is true that their appearance can vary according to the algorithms used to render the image. But the topology of the network, which is a hyper-dimensional object, is invariant to how it is rendered in two-dimensional space – just like a three-dimensional building, which can be photographed from many angles, remains the same regardless of how it is captured. Conclusions and analyses do not rely on the visual appearance of a network. And, there are highly developed techniques of diverse sorts for "optimally" rendering a network in two dimensions, which we exploit [57, 58].

In *Connected*, we call the empirical regularity that clusters of behaviors or attributes extend to three degrees of separation the 'three degrees of influence rule' [27]. We realize that this telegraphic phrase can be seen as problematic by some. For instance, so far, the evidence offered above pertains to clustering, not influence; moreover, the use of the word 'rule' may imply a degree of determinism that is too strong. However, similar to the widespread use of the expression 'six degrees of separation,' this turn of phrase is meant to be evocative, not definitive. For instance, even the widely discussed 'six degrees of separation' is not precisely six, neither in Milgram's classic paper [59] nor Watts and colleagues' clever, well known email experiment [60].

Our objective – insofar as influence is concerned – is to make the point that (1) there is evidence that diverse phenomena spread beyond one degree, and (2) there is evidence that the association fades within a few degrees in what seems like a systematic way across phenomena and datasets. Incidentally, work by other investigators on the spread of ideas (such as where to find a good piano teacher or what information inventors are aware of) similarly seems to identify an important boundary at three degrees at which meaningful effects are no longer detectable [61, 62]. And recent work using large twitter datasets also confirm the clustering of happiness (as measured using text processing) to three degrees of separation [63]. Finally, the boundary at three degrees does not need to be sharp, nor would it be expected to be; rather, as discussed below, this empirical regularity probably reflects the point at which effects are simply no longer statistically discernable even with reasonably large datasets.

Regarding the role that interpersonal influence plays in clustering to three degrees of separation, we frequently make the point that different things spread in different ways and to different extents. Hence, we also find that the actual number of degrees of separation at which any clustering is (statistically) detectable, and at which any spread is therefore likely, varies depending on the behavior or the observational or experimental context. For instance, Figure 1 shows that, using diverse data sets, we have found evidence of clustering (and hence, possibly, of spread) to two degrees of separation (divorce) [23] and four degrees (drug use, sleep) [22]. Moreover, we have found evidence of spreading in the laboratory as well; in an experimental study of cooperation in public goods games (with full ascertainment of ties and no threat from homophily or confounding), we found that behavior spread to three degrees [6]. Whether *three* ends up being the modal pattern remains to be seen. But we do not think that the value itself is the issue. It is the fact that it is greater than *one* that really interests us. Moreover, and on the other hand, it's not too great either: if a given person's actions could indeed spread to six degrees of separation, what we know about the connectedness of people on the planet would suggest a kind of global influence of a single individual that seems very implausible.[5]

In most of our papers, we use regression methods to discern whether there is evidence for person-to-person spread, and these methods often suggest that things do *not* spread. For example, in our obesity paper, we find evidence of correlation between friends but not between neighbors (see Figure 4 here). Moreover, some things, like health screening behavior [18] and sexual orientation [22], do not appear to spread across any observed social ties in our analyses. This is noteworthy because we have never claimed that *everything* spreads, and the same methods that have been used to develop evidence of spread in some phenomena fail to show spread in other phenomena.

Not only is it the case that not everything spreads, but it is also the case that not everything spreads by the same mechanism. For example, weight gain may spread via imitation of a specific eating behavior (e.g., eating fried foods), imitation of a specific exercise behavior (e.g., jogging), or adoption of a social norm that yields changes in overall behavior. If it is the norm that is transmitted, then other specific behaviors may not be correlated: a person who starts jogging may influence his friend to take up swimming or reduce eating, and both individuals may lose weight as a result.

Interestingly, the permutation results raise the possibility that the spread of traits may skip over a person in a given chain. If the only way something like obesity spreads is via realization of a change at each step on the path between two individuals, and if there are only three individuals connected by two social ties (i.e., if there is only one path – we discuss this assumption in the next section), then the probability that a person affects his friend's friend should be the square of the probability that he affects his friend. If Joe has a 20% chance of influencing John, and John has a 20% chance of influencing Mary, then Joe should have a 4% chance of influencing Mary (if we assume that the probabilities are independent). But that is not what we find. The associations in traits do not decay exponentially. As a consequence, it may be the case that some people can act as "carriers" who transmit a trait without exhibiting it themselves (similar to certain pathogens). For example, a person whose friend becomes obese may become more accepting of weight gain and as a consequence may stop encouraging *other* friends to lose weight even if his own weight does not change. Such latent transmission is additive to the manifest transmission. This is one possible explanation, among others, of why the effects observed are not simply or exactly multiplicative.

There are at least two explanations for the apparent limit at roughly three degrees (we discuss others in our book, *Connected*). The first and simplest is decay, or a decline in size of meaningful or detectable effects. Like waves spreading out from a stone dropped into a still pond, the influence we have on others may eventually just peter out. In social networks, we can think of this as a kind of social "friction." Of course, ascertaining decay depends in some sense on the sample size and the effect size. With big samples and/or big effects (and with complete network ascertainment), any clustering that extends to further distances – even if unimportant – could be detected. In short, the empirical regularity of three degrees of influence may simply reflect a decay in the size of an effect to the point were the effect is no longer detectable.

Second, influence may decline because of an unavoidable evolution in the network that makes the links beyond three degrees unstable. Ties in networks do not last forever. Friends stop being friends. Neighbors move. Spouses divorce. People die. The only way to lose a direct connection to someone you know (geodesic distance 1) is if the tie directly between you disappears. But for a person three degrees removed from you along a (solitary) simple chain, any of three ties could be cut and you would lose the connection. Hence, on average, we may not have stable ties to people at four degrees of separation given the constant turnover in nodes and ties all along the way. Consequently, we might not influence nor be influenced by people at four degrees and beyond. The extent to which such an effect holds empirically, however, will also depend on the nature and number of redundant paths between people at various degrees of separation, as described below.

---

[5] Connectivity (either at six degrees or any other geodesic distance) sets an upper bound on influence. Moreover, it is worth emphasizing that three degrees (plus or minus one degree) is actually a lot smaller than six, because the number of paths grows exponentially (or even faster) as a function of geodesic distance.

### (3) Partial Observation of FHS-Net Ties

Some commentators have expressed concern that our findings related to clustering to three degrees of separation might relate to the nature of sampling in the FHS-Net. In particular, subjects only name a limited number of friends (generally only one person at any given time, a person who can be thought of as the subject's one "best friend"), which leaves open the possibility of unobserved "backdoor" paths between nodes. The concern is that if nodes or edges are not observed, then two individuals who are actually one or two degrees apart might be wrongly supposed to be three (or more) degrees apart. Stated another way, the claim that a person's traits are related to the same traits of a person three degrees removed from them might be false because a partially observed network might miss pathways that would otherwise show these individuals to actually be only one or two degrees removed.

This is a sensible concern. However, the intuition that partial observation will necessarily lead to overestimation of the length of the path over which influence is transmitted is incorrect. First, it is important to distinguish between three types of paths: (1) the actual, inherently unobservable, stochastic path taken by the germ, norm, or behavior that spreads; (2) the shortest path, and hence the most likely single actual path, between the source and target nodes in the fully observed network; and (3) the shortest path(s) between the source and target nodes in a partially observed network. This is illustrated in Figure 3. Although the actual paths cannot be observed in practice, one can nevertheless explore the relationships between these three path lengths using simulations.

Extensive exploration of a network of 3.9 million cell phone users and the ties between them, as captured by their call records, reveals that, counter-intuitively, the shortest paths in a sampled (observed) network may be shorter than the actual paths [11]. In other words, when specific paths of varying lengths taken by a diffusion process exist between pairs of individuals within a network, and when these paths are sampled, it turns out that the sampled path lengths can be shorter *or* longer than the actual paths. The specific outcome depends on the extent of sampling of nodes and ties, but the actual paths are typically roughly 10% - 30% shorter than the shortest paths in the partially observed networks for many sampling frames. Consequently, the intuition that partial observation will necessarily lead to an inflation in measured path length (and hence possibly to a mis-measurement of clustering) is incorrect.

The reason for this is as follows. Imagine that the shortest path in the non-sampled (fully observed) network connecting the source and the target nodes has a length of, say, three steps. We would take this as the most probable path of spread of some phenomenon. Now imagine that, because of the sampling process, part of this path vanishes (i.e., we can no longer observe it). Following the same logic that the most probable infection path between the source and the target nodes is the shortest path connecting them, we now find the shortest path in the sampled (partially observed) network between the source and the target nodes. This cannot be shorter than three, but it may be equal to three if there were multiple paths of that length, and it may also be longer than three.

Suppose that the shortest observed path has a length of four. Although the shortest path is the single most likely path between the two nodes, it is not the only path between them. Depending on the structure of the network, there may be multiple paths of length four, and although each of them taken separately is less likely to be observed than the path of length three, the overall probability that the transmission happens through four steps versus three steps depends on the number of paths of these lengths. In real human networks, it is frequently the case that once we let a spreading front proceed a few steps from the source, the length of the actual path between the source and target nodes is higher than the shortest length. If that were the case in our example, detecting the shortest path of length three in the fully observed network would lead to an under-estimate of the actual path. Because partial observation may inflate our estimate of the shortest paths, it may hence, counter-intuitively, reduce the net bias of the estimated length of the actual path.

Furthermore, equally important with respect to the concern regarding partial observation, we find similar clustering, to three degrees of separation, in data sets where networks ties were almost *fully* observed, as shown in Figure 1. For example, in the National Longitudinal Study of Adolescent Health, subjects were asked to name up to 10 friends, and 90% of them named fewer than the maximum. And, in a paper on the spread of sleep behavior and drug use in this particular network, we actually find clustering up to four degrees of separation [22].

### (4) Basic Analysis and Findings: Longitudinal Regression Models

The topological permutation tests described above test only simple null hypotheses of no association (albeit in a way that permits more than dyadic ties). In order to explore the possible reasons for the clustering described above, we studied more closely the person-to-person relationship using a regression framework. We specified longitudinal regression models with a basic form wherein the ego's status (e.g., obese or not) at time $t+1$, denoted $y_{t+1}^{ego}$ (with distribution $Y_{t+1}^{ego}$), was a function of various time-invariant attributes of egos, such as gender and education (captured by the $k$ variables denoted by $x$ on the right), their status at time $t$ ($y_t^{ego}$), and, most pertinently, the status of their alters at times $t$ ($y_t^{alter}$) and $t+1$ ($y_{t+1}^{alter}$).[6] This model could be generalized to allow for time-varying control variables of the ego, and to allow for attributes of the alter to be included as well.

We used generalized estimating equations (GEE) to account for multiple observations of the same ego across waves and across ego-alter pairings [64]. And we only included observations in which ego and alter had a relationship at both time $t$ and time $t+1$ – on the grounds that people who are disconnected from each other should not influence each other that much, if at all (though this is a constraint that can – informatively – be relaxed) [65].[7] In general, interpersonal ties within the FHS-Net were very stable [9].

---

[6] These models are similar to models described by Valente [51].

[7] Here, this kind of "disconnection" is different than another kind: people can be disconnected from each other (in the sense that there is no path at all between them through the network) or they can be disconnected in that they have

Our basic model is thus:

$$g\left(E\left[Y_{t+1}^{ego}\right]\right) = \alpha + \beta_1 y_t^{ego} + \beta_2 y_{t+1}^{alter} + \beta_3 y_t^{alter} + \sum_{j=1}^{k} \gamma_j x_j \qquad (1)$$

where $g()$ is a link function determined by the form of the dependent variable. For continuous data, $g(a)=a$ and for dichotomous data, $g(a)=\log(a/(1-a))$. In most of our articles, we specify both link functions, for instance, studying dichotomous obesity and continuous BMI, or studying dichotomous heavy smoking and continuous measures of how many cigarettes per day a person smokes.

Since we are using GEE, we also estimate the covariance structure of correlated observations for each ego. The covariance matrix of $Y^{ego}$ is modeled by $V_{ego} = \phi A_{ego}^{1/2} R A_{ego}^{1/2}$ where $\phi$ is a scaling constant, $A$ is a diagonal matrix of scaling functions, and $R$ is the working correlation matrix. We assumed an independence working correlation structure for the clustered errors, which has been shown to yield asymptotically unbiased and consistent, although possibly inefficient, parameter estimates (the $\beta$ and $\gamma$ terms) even when the correlation structure is mis-specified [66].

To be clear, our basic model assumes that there is no correlation of ego's weight at $t+1$ with alter's weight at $t+1$ except via influence, and no other effects on ego's weight at $t+1$ except via the effect of ego's past weight at time $t$ and the effect of the measured covariates, i.e., conditional on no unobserved confounding. These are common assumptions in regression models of observational data, of course. However, a special consideration here is that this assumption implies that there is no unobserved homophily beyond that on the observable variables. Moreover, pertinently, these models are specified for each alter type independently (unless comparisons between types of alters are sought, in which case one could, for instance, specify a 'sibling' model and index the kind of siblings at issue).

The time-lagged dependent variable (lagged to the prior exam) typically eliminates serial correlation in the errors when there are more than two time periods observed in the case of AR(1) models in which the Markov assumption holds. We test for significant serial correlation in the error terms using a Lagrange multiplier test [67], and, in all cases we have studied, the correlation ceases to be significant with the addition of a single lagged dependent variable. Inclusion of this variable also helps control for ego's genetic endowment or any intrinsic, stable predilection to evince a particular trait.

The lagged independent variable for an alter's trait helps account for homophily (especially with respect to the observed trait that is the object of inquiry) because it makes ego's current state unconditional on the state the alter was in when the ego and alter formed a connection [51]. Conditioning on the lagged alter's trait, however, would not comprehensively deal with homophily on *unobserved* traits that are both time-varying and also associated with the outcome of interest (for instance, if people who unobservably knew they wanted to lose weight preferentially formed ties with other similar people). This term also does not address the issue of a shared context (confounding).

Note also that our base model can be regarded as an equation expressing the effect of alter's baseline weight and alter's *change* in weight. The generative interpretation is that the control for alter's and ego's baseline weight controls for homophily on weight, and the other terms address the impact of a change in weight. Thus, our model is closely related to auto-

---

no direct connection (and have only an indirect connection – e.g., they are a friend's friend). In the latter case, as argued here, they can affect each other via a sequence of dyadic ties.

distributed lag (ADL) and error correction models (ECM) that are frequently used in time-series econometrics to evaluate the extent to which two series that tend towards an equilibrium coupling covary [68]. In particular, one can think of the coefficient on the contemporaneous alter characteristic as a measure of the "short run" or one-period effect of the independent variable on the dependent variable (in this case, of the alter on the ego) [69]. According to this interpretation, an alter may experience a shock to some attribute (they may gain 10 pounds, for example) and the coefficient on alter weight would then tell us the size of the first change back towards the equilibrium coupling of ego and alter weight. Figure 4 illustrates some of the results we have published using such longitudinal models.

Importantly, we also specify models with further lags in the alter variables in most of our work, evaluating how the change in a trait in an alter between t-1 and t is associated with the change in a trait in an ego between t and t+1. Although these models are underpowered compared to the approach we describe above, they typically suggest comparable positive effect sizes. As noted by VanderWeele, Ogburn, and Tchetgen [70], this approach effectively responds to many of the concerns articulated by one critic [49], including any concerns about model consistency or test validity.

We have tried to be clear about other assumptions underlying our technical specifications both here and in our prior published work. For example, we do not believe it is necessary to specify a single, joint model for all the effects present. Notably, in our exploration of various datasets, we sometimes interact key variables with the relationship type, but these models have never suggested that we would arrive at different conclusions by modeling multiple relationship types at the same time in a single dyadic model. Moreover, whether particular assumptions are required for model estimates to be taken as identifiable will often depend on the eye of the beholder – for instance, whether it is plausible to assume that there is no meaningful homophily on unmeasured traits that also affect the trait of interest. Even given such constraints and restrictions, however, we believe that the results of such modeling exercises are of interest; moreover, they give other scholars an opportunity to explore how the estimates change when variables are added to the model or model assumptions are relaxed.

We also note that there are certainly other valuable ways of analyzing such data, albeit with other strengths and limitations (such as constraints on network size and on parameter interpretation), including the so-called "actor-oriented" models [71, 72]; see [73, 74] for illustrative applications. These models also involve their own assumptions, of course, and these models do not escape some of the general criticisms of the use of observational data, despite any claims to the contrary.[8] In our case, we did not use this approach, ably described by Snijders and colleagues, because our sample sizes were bigger than the models could accommodate. And, of course, there is a long-standing appreciation of the difficulty of causal inference of peer effects, for which an early and lucid articulation was provided by Charles Manski [76].

---

[8] For example, Lewis, Gonzalez, and Kaufman [75] claim that SIENA models suffer less from the threats to causal inference posed by observational data. However, SIENA is susceptible to contextual effects and indirect homophily just like any other statistical model of observational data. Moreover, their study has a number of other noteworthy limitations that subvert the plausibility of its conclusions, including, (1) it treats "weak tie" Facebook friends the same as the "strong tie" real friends among a person's Facebook friends; it should have been expected that tenuous ties to acquaintances would not evince much influence; (2) it starts with 1,600 people, but only analyzes 200 for whom they have complete data, and the analyses do not account well for this missingness; (3) perhaps because they have only 200 cases, their confidence intervals are wide, though the point estimates for inter-personal influence are actually typically large; (4) it reports that many of their models did not converge (a problem that plagues SIENA); (5) there is no evidence that the models they report converged (they do not report any convergence diagnostics such as the Raftery-Lewis test). In contrast, a study of ours involving a randomized controlled trial of 61,000,000 people in Facebook shows significant levels of inter-personal influence online [3].

Our basic modeling framework has attracted some specific criticisms about the extent to which homophily and confounding can indeed be purged from the causal estimates [15], about whether this model is capable of offering any insight into the effects at hand at all [77], or about the nature of various biases that might by introduced by changes in network topology across time [78].[9] In most cases, we discussed these potential limitations in our original papers. We have also previously published two discussions of some of these concerns [15, 79], and we describe some of the other, newer issues below. We recognize the valuable contribution that these critiques have made to advancing the field of estimating network effects using observational data.

But it is also fair to say that these critiques in some cases simply re-state the generic claim that it is difficult (some say impossible) to extract causal inferences from observational data at all. But here we do not engage this essentially nihilistic position: it is not specific to our own work or even to the issue of causal inference using network data, and so it is well beyond our present scope.[10] Another paper, among other things, basically asserts that any modeling of observational data is suspect since such modeling must rely on assumptions [49]; not only do we reject this nihilistic position as well, but the claims of this author have either been retracted by the author (in an erratum published after the fact) or substantially refuted by others [70]. In short, we believe that the key issue is the extent to which one can be explicit about one's assumptions, and the reasonableness of those assumptions, in work analyzing social networks as in any other statistical work.

One paper that attempted to refute the regression approach embodied in equation (1) claimed to document the spread of phenomena in adolescents which were assumed to be intrinsically incapable of spread, such as acne, headaches, and height [81]. Interestingly, and in accord with this assumption, the authors indeed find no effect for the first two outcomes (acne and headaches), at conventional levels of statistical significance ($p$=0.05). But they stretch the threshold to $p$=0.10 so they can make the claim that these outcomes do spread, even in a dataset that is large (over 5,000 people). Then they make their argument: since these outcomes could not possibly spread, the regression framework must necessarily somehow be intrinsically wrong.

It is worth noting, however, that, in addition to not being statistically significant at conventional levels, the effect sizes for these phenomena were also small, substantially smaller than the effects observed, for example, for obesity and smoking in the networks we have studied, including both the FHS-Net and Add Health data. Indeed, these effects (for acne, headaches, and height) are not robust to sensitivity analyses for the role of homophily or shared context, as shown by formal sensitivity analyses conducted by others [82].

---

[9] We note, however, that the levels of change in friendship seen in the FHS-Net (as documented in O'Malley and Christakis [9]) are sufficiently modest that they would not be consistent with much bias of the kind suggested by Noel and Nyhan in any case (even judging from the estimates in the Noel and Nyhan paper).

[10] We are of the opinion, however, that the world is knowable and that careful observation of the world has a very important role to play in knowing it, and even that it is indeed possible to make causal inferences from observational data. One of our favorite illustrations of this is that we know that jumping out of a plane is deadly, even though there has never been a randomized trial of this 'treatment.' One tongue-in-cheek paper that attempted to do a meta-analysis of use concluded: "As with many interventions intended to prevent ill health, the effectiveness of parachutes has not been subjected to rigorous evaluation by using randomised controlled trials. Advocates of evidence-based medicine have criticised the adoption of interventions evaluated by using only observational data. We think that everyone might benefit if the most radical protagonists of evidence-based medicine organised and participated in a double blind, randomised, placebo controlled, crossover trial of the parachute" [80].

Moreover, it is in fact *not* inconceivable that such small contagion effects for acne, headaches, and even height (in adolescents) might indeed exist. First, it must be remembered that, unlike the FHS-Net, the dataset used in these analyses (Add Health) captures only *self-reported* outcomes. Hence, if an ego has a friend who complains of headaches, the ego might find it easier to complain of headaches (either because he has been given license to, or because he finds it normative). Conversely, perhaps the ego's friend has discovered an effective means to treat headaches and has communicated it to the ego, and so both ego and alter might take medication for headaches, thus explaining the diffusion of the presence or absence of headaches [83]. As for acne, whether an ego deems the few pimples on her face to be worthy of report as "acne" may be influenced by her friend's perceptions of her problem or her friend's appearance or what her friend has told her that she should think about these pimples. Or, the friend might encourage the ego (or show her how) to treat the acne, such that the ego's acne status does indeed come to be influenced by the friend's.

At first pass, it would seem that height should not spread. Yet, in adolescents, it is not inconceivable that it might, and environmental factors explain a significant portion of the variance in height (around 20%) prior to adulthood [84]. To the extent that adolescent growth is, as is well known in the medical literature, influenced by exercise, nutrition, and smoking, it is entirely possible that an adolescent's height could depend (to some degree) on the height of his friends, to the extent that they share smoking or exercise habits, for example. Moreover, adolescents with tall friends could (and perhaps would) again *report* that they are taller than they really are, or that they were gaining height faster than they really were – since, unlike the FHS-Net where height was measured by nurses, in AddHealth, it was self-reported. Hence, overall, like the spread of obesity, it is not *literally* the obesity or height that spreads, but norms and behaviors (e.g., about exercise, nutrition, smoking) that do. These could induce a correlation in height gain between friends that is not induced between strangers.

Thus, from our perspective, even if the authors had shown that all three phenomena (acne, headaches, and height) spread among growing adolescents at conventional levels of significance, this would not have been a fatal blow to the statistical methods that they are criticizing let alone to the claim that health phenomena can spread.[11] And, again, our own work with various outcomes in this modeling framework has often yielded results that show that phenomena do not spread.

Moreover, a recent paper by VanderWeele is very informative [82]. He applied sensitivity analysis techniques [85] to several of our papers, as well as to some analyses conducted by others. In particular, he estimated how large the effect of unobserved factors would have to be in order to subvert confidence in the results. He concluded that (subject to certain assumptions) "contagion effects for obesity and smoking cessation are reasonably robust to possible latent homophily or environmental confounding; those for happiness and loneliness are somewhat less so. Supposed effects for height, acne, and headaches are all easily explained away by latent homophily and confounding."

---

[11] The further claim by these authors that adding additional controls for environmental factors attenuates the effect is also limited, and we examined this possibility in the FHS-Net in a variety of ways. It is important to note that friends in Add Health are all physically proximate (they are in the same school), whereas this is not necessarily the case in the FHS-Net. If our estimates are biased because they capture community-level correlation, one implication is that the increased geographic distance between friends will reduce the effect size (since distant social contacts are not contemporaneously affected by community-level variables). But, as noted below, we find that the relationship does *not* decay with physical distance, even up to hundreds of miles away.

This does not mean, of course, that the modeling framework of equation (1) is in fact free of any bias or is perfectly able to capture causal effects. This is one of the reasons we described exactly what models we implemented both here and in our published papers and their supplements, as well as additional innovations that we attempted within this framework, such as a novel identification strategy exploiting tie direction.

## (5) An Identification Strategy Involving Directional Ties

In our first paper, we proposed an identification strategy that we thought could provide additional evidence regarding the causal nature of peer effects. Just as researchers use the directional nature of *time* to establish a sequence that is consistent with a causal ordering, we tried to use the directional nature of *ties* to do the same. Specifically, we suggested that differences in effects according to the asymmetric nature of social ties could shed light on the possibility of confounding due to extraneous factors[14].

A key element of sociocentric network studies involving friends is that all subjects in the specified population identify their social contacts. As a result, we have two pieces of information about every friendship: (1) whether the ego nominates the alter as a friend, and (2) whether the alter nominates the ego as a friend. Because these friendship identifications (unlike, say, spouse or sibling ties) are directional, we can study three different kinds: an "ego-perceived friend" wherein ego nominates alter but not vice versa; an "alter-perceived friend" wherein alter nominates ego but not vice versa; and a "mutual friend" in which the nomination is reciprocal. We theorized that the social influence that an alter has on an ego would be affected by the type of friendship we observe, with the strongest effects occurring between mutual friends, followed by ego-perceived friends, followed by alter-perceived friends. [12] For alter-perceived friends, we might even expect *no* effect at all, since ego might not be aware of alter, let alone alter's actions. The model in equation (1) can be specified for different sorts of ego-alter pairings including different "kinds" of friends. Of course, in the case of friendship, these models can be specified for friends in general and an indicator can be added to index the kind of friendship.

Figure 4 shows that this pattern of results generally exists for a wide range of behaviors and affective states – in two different data sets. Evidence regarding the directional nature of the friendship effects is important because it suggests that covariance in traits between friends is unlikely to be the result of unobserved contemporaneous exposures experienced by the two persons in a friendship. If it were, there should be an equally strong association, regardless of the directionality of friendship nomination. We also proposed a similar argument, using just triads of people (specifically, men paired with their wives and ex-wives) in another paper [65]. And, in another recent paper, we lay a foundation that could allow the use of an asymmetry in ties that is continuous rather than dichotomous [10].

One commentator has asserted that we have somehow misrepresented these results [49]. In most cases, the confidence intervals for the three types of friendships overlap; in our papers, we have noted the ordering of the effects and reported their confidence intervals so as to evaluate the directionality pattern. And all explicit claims of significance that address pairwise differences in point estimates have contained confidence intervals or *p*-values for the comparison (derived from a single model with an interaction term).

---

[12] A paper by Mercken et al [86] highlights the socially more important role of reciprocated friendships compared to unreciprocated friendships but does not pursue this difference as an identification strategy.

Moreover, it is important to note that answering the question about whether or not the pattern (mutual tie > ego-perceived tie > alter-perceived tie) is true (i.e., can be stated with confidence) depends on the null hypothesis. For example, what is the probability that all three of the different kinds of relationships are drawn from the same distribution? How likely is the order of the effects to be as specified? Such considerations would give a different result than a test of whether or not two of them were drawn from the same distribution. And what is the likelihood that we would find this ordering over and over again, including in different network data sets (as shown in Figure 4)?

The strengths and limitations of this network directionality test have since been explored by computer scientists [87], econometricians [88], statisticians [77], and others [78]. Possibly, there are papers from before 2007 exploiting the directionality of ties as well, of which we are unaware. One paper in particular identifies two further, important assumptions that may be necessary or implicit in the directional test [77]. Specifically, they argue that if two conditions are met, the test becomes less reliable as a way to exclude confounding. These two conditions are (1) the influencers in a population differ substantially and systematically in the unobserved attributes (X) from the influenced in a population, and also that (2) the different neighborhoods of X have substantially different local relationships to Y (the outcome). How likely such circumstances are to occur in real social networks is unclear, and how big any resulting biases might be is also unclear; again, like so many discussions of statistical methods, the utility of the method critically hinges on the question of what assumptions are "reasonable." We believe that the foregoing circumstances do not realistically hold to a large extent, at least in general, given what is known about social systems.

Finally, a recent paper by Iwashyna et al. [89] uses agent-based models to generate network data with varying processes of friend selection and influence. The authors then perform a GEE regression analysis like the one we have used to measure its sensitivity and specificity in detecting influence and homophily in data where the underlying processes are known. They show that the model works well to detect influence, with a very high sensitivity and high specificity, but that it does not work well to detect homophily. A particularly important feature of their work is that it addresses the "latent homophily" argument made by Shalizi and Thomas (2010) who argue that covariates that affect both homophily and the outcome can bias the model (though Shalizi and Thomas do not quantify this bias). Iwashyna et al. actually test a specification where people make friends based on an unobservable characteristic related to the outcome, and yet they still find the GEE model for inference yields high sensitivity and specificity for detecting influence. Thus, while there may be some theoretical objections based on unknown amounts of bias that could be present in our results, applied research is generally pointing to the utility of the approach in generating informative estimates of the possible inter-personal influence present.

(6) Using Geographic Information to Address Certain Types of Confounding

Another important advantage of the FHS-Net is that, in addition to the social network, we also have information about place of residence (including as it changes across time). This means that we can calculate not only social distance, but also geographic distance, between any two people. And since participants in the FHS-Net have spread out over the country, there is substantial variation in this measure; ordered by distance between friends' residences, the last sextile of the distribution averages nearly 500 miles. This is important because it helps us to discern whether or not changes in the local physical or social context might explain the correlation in outcomes between two people who have a relationship. For example, the opening of a popular fast food restaurant might cause many people in an area to gain weight, and this contextual effect might cause us to falsely infer that peers are influencing one another.[13]

Instead, we found in the obesity paper and in other follow-up studies of health behavior (smoking, drinking) that distance played no discernable role in the correlation in outcomes. An interaction term between geographic distance and alter's outcome at time $t+1$ yields a coefficient that is near 0 and insignificant. In other words, a friend who lives hundreds of miles away appears to have a similar effect as a friend who lives next door. Social distance appears to matter much more than physical distance. Since these models, as before, condition on the lagged trait value for the egos and the alters, homophily on the trait of interest is also an unlikely explanation.

On the other hand, when we turned to studies of affective states (happiness, loneliness, and depression) we found a different result. Associations were only positive for friends and siblings who lived nearby (within a few miles). One interpretation of this result is that affective states require physical proximity to spread, and this would be consistent with the psychological literature on the spread of emotions via face-to-face contact [27]. But it is also possible that these results are being driven by contextual effects: people in a given neighborhood, exposed to the same environment, might tend to react by changing moods in the same direction to the same stimuli. To evaluate this possibility, we compared the associations in outcomes for *next-door* neighbors to those for *same-block* neighbors (people who live within 100 meters of one another). And although we had many more observations at the block level, the association in outcomes was significant for the next-door neighbors and not for others. Thus, while it is still possible that contextual effects explain some of the association, they would need to be "micro-environmental" contextual effects that would not affect everyone on the same block.

### (7) Availability of Data and Code

Some commentators have asked about data availability. We have developed and placed into the public domain much network data and code (including for the Facebook network [2], biological networks [91], experimental networks [6], and various political network datasets [92-95], and we have promptly shared our code and supplementary results with anyone who has asked (e.g., [78]). Of course, the Add Health is a publicly available dataset, so anyone wishing to explore new analytic approaches to network data, or the assumptions required to analyze such data, may take advantage of it. There are many other sources of social network data as well (e.g., online data), though longitudinal data are still somewhat scarce.

With respect to the FHS-Net, we worked closely with FHS administrators to release the data. Regrettably, given the origin of these data in clinical records and given FHS rules, not all the data was releasable, which affects the replicability of our results (at least those results with FHS data) by outside researchers. However, we have shared data with collaborators using our secure servers. And, in 2009, the study's administrators, with our assistance, posted a version of these data in a secure online NIH repository that requires formal application procedures. FHS

---

[13] Interestingly, a careful analysis of the FHS reveals no effect of proximity to fast food establishments, so this example is just hypothetical; see Block et al [90]

implemented a variety of changes to the data in order to help protect subject confidentiality, however, before posting. Specifically: (1) all date information was changed to a monthly resolution rather than daily; (2) only 9,000 cases rather than 12,000 could be posted (e.g., all non-genetically related relative ties, such as adopted siblings, step-children, etc., were removed); (3) individuals who did not consent to the release of 'sensitive information' were excluded; and (4) the available covariates (e.g., geographic coordinates) were restricted. We have re-run some of our analyses on this restricted dataset, and many – but not all – of our results survive these restrictions. This dataset is distributed via the SHARE database at dbGAP (http://www.ncbi.nlm.nih.gov/gap).

<u>(8) Social Influence and Social Networks</u>

We believe that we have been careful in interpreting our findings and that we have summarized our results with the proper caveats. For instance, the first two paragraphs of the Discussion in our 2007 paper on obesity read as follows:

> "Our study suggests that obesity may spread in social networks in a quantifiable and discernable pattern that depends on the nature of social ties. Moreover, social distance appears to be more important than geographic distance within these networks. Although connected persons might share an exposure to common environmental factors, the experience of simultaneous events, or other common features (e.g., genes) that cause them to gain or lose weight simultaneously, our observations suggest an important role for a process involving the induction and person-to-person spread of obesity.
> "Our findings that the weight gain of immediate neighbors did not affect the chance of weight gain in egos and that geographic distance did not modify the effect for other types of alters (e.g., friends or siblings) helps rule out common exposure to local environmental factors as an explanation for our observations. Our models also controlled for an ego's previous weight status; this helps to account for sources of confounding that are stable over time (e.g., childhood experiences or genetic endowment). In addition, the control in our models for an alter's previous weight status accounts for a possible tendency of obese people to form ties among themselves. Finally, the findings regarding the directional nature of the effects of friendships are especially important with regard to the interpersonal induction of obesity because they suggest that friends do not simultaneously become obese as a result of contemporaneous exposures to unobserved factors. If the friends did become obese at the same time, any such exposures should have an equally strong influence regardless of the directionality of friendship. This observation also points to the specifically social nature of these associations, since the asymmetry in the process may arise from the fact that the person who identifies another person as a friend esteems the other person" [14].

We stand behind this summary.

Some who have found fault with our analyses or conclusions have seemed, in reality, to find fault with second-hand accounts of the work. One of the more frustrating experiences we have had is to be criticized for overlooking limitations in our data or methods that we did not, in fact, overlook, but that were instead overlooked by others who were describing or summarizing our work (often for a lay audience). In reality, we carefully laid out and explored nearly all of these limitations in our published research and our public presentations to scientific audiences. While we have sometimes speculated about mechanisms of inter-personal effects, we have avoided making strong mechanistic claims in our scientific papers (though we have been a bit more willing to hypothesize in *Connected*, intended for a non-scientific audience).

Our work depends, of course, on many who came before us, and there is a long tradition of looking at peer effects in all sorts of phenomena, particularly in dyadic settings. Our writings cite prior work by many other scientists. Moreover, since we published our work, a variety of articles by other investigators have used other data sets and approaches and confirmed the findings and, in many cases, even the magnitude of the effects we observed. Pertinent recent work with obesity, weight gain, weight loss, and the mechanisms and behaviors related to this (e.g., eating, exercise) that mostly confirm our findings is quite diverse, including everything from observational studies, to natural experiments, to *de novo* experiments, to twin studies that account for genetic similarity, to clever studies involving electronic monitoring of interactions [24, 96-109]. One experimental study documented the spread of weight loss across spousal connections; the spouses of individuals randomly assigned to weight loss interventions were tracked, and evidence of a ripple effect was apparent from the subjects to their (untreated) spouses [53]. Of course, much work, as expected, has also confirmed the existence of homophily with respect to weight (e.g., [9, 110]). Still other studies have used experimental and observational methods to confirm the idea that one mechanism of inter-personal spread of obesity might be a spread of norms, as we speculated in our 2007 paper (e.g., [102, 107, 111, 112]).

There is also a longstanding literature on emotional contagion, of course [113], but recent social network papers that have confirmed our findings have also appeared [114-116]. Other outcomes have also recently received a re-evaluation, such as smoking (which, of course, has its own longstanding literature with respect to peer effects) [117-119], with many papers identifying the obvious importance of both homophily and influence, especially in adolescent populations (see, e.g., [73, 74]). Indeed, there have been a number of randomized controlled trials of smoking cessation interventions that target students based on their network position and that documented peer effects, an approach that was thoughtfully pioneered by Valente and colleagues [120-122].

A key consideration, therefore, is what the standard for evaluating our findings is. Is the real issue whether such interpersonal influence for these interesting phenomena (obesity, emotions, etc.) occurs? In that case, confirmatory work of various types by various investigators should be taken to support our findings. Here, the standard is whether an observation is true or not. In this regard, we think the body of evidence accumulated about peer effects – if not network effects – is very persuasive, and we are joined in this view by many social and biomedical scientists.

Or is the key issue here that interpersonal effects are hard to discern with confidence, and that data and methods are imperfect and subject to assumptions or biases? If so, we quite agree. This is one of the reasons we have tried to be transparent about the methods used in our work. This is also one of the reasons that we ourselves, and others working collaboratively with us, have proposed new approaches, such as experiments (both offline and online) [3, 6, 7, 123],[14] and instrumental variable methods involving genes as instruments [125], both of which might be able to provide different sorts of confidence in causal inference. Here, the standard is whether an accurate observation is scientifically possible. We think it is. Since network data are likely to become increasingly available in this era of computational social science [126], and since questions regarding the structure and function of social networks are of intrinsic importance, it seems clear that innovation in statistical methods will be required. We are eager to hear of any practical approaches to the analysis of large-scale, observational social network data that shed additional light on the interesting and important phenomenon of inter-personal influence.

---

[14] Sinan Aral and colleagues have also been advancing this area; see, for example, Aral and Walker [124].

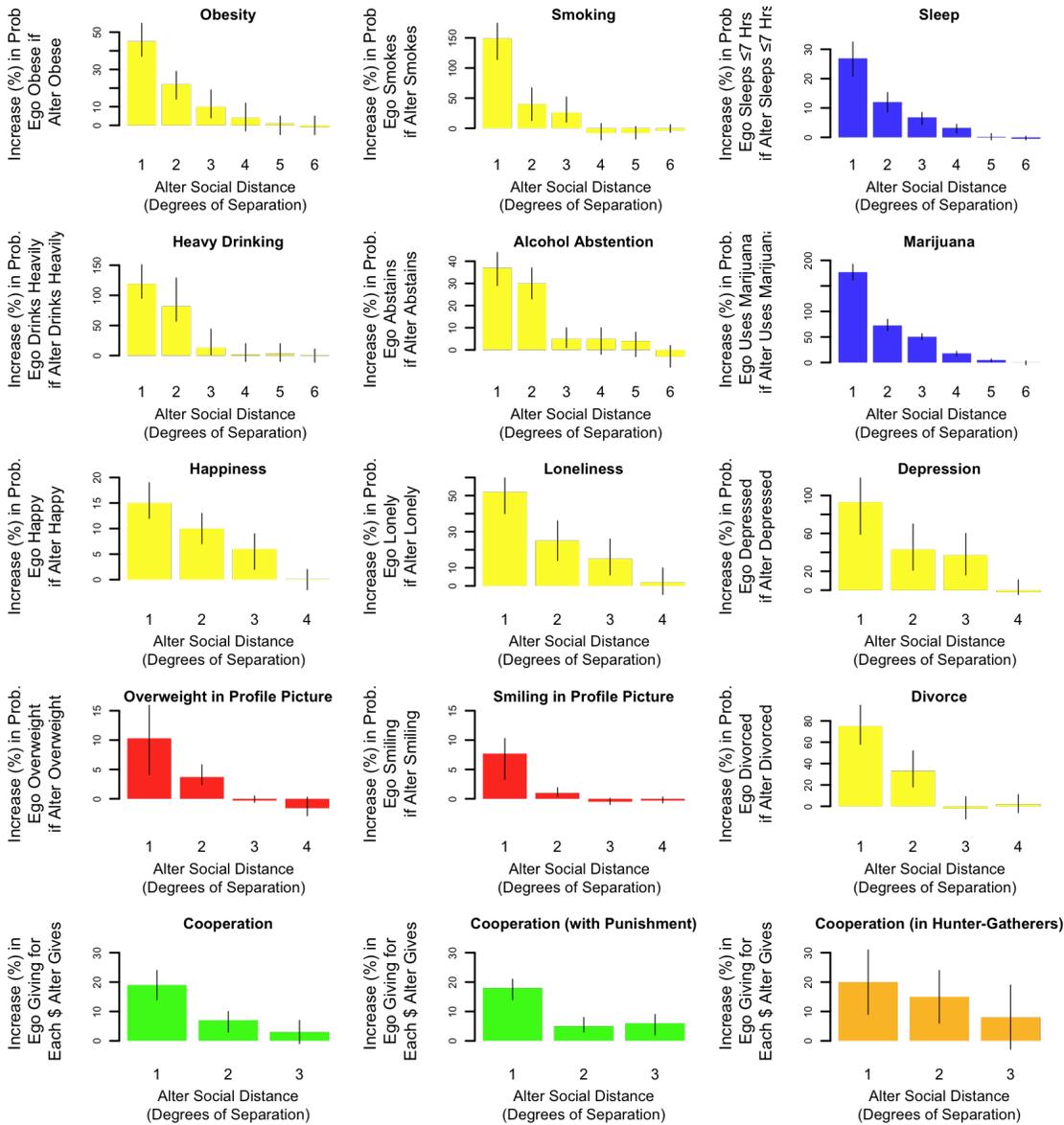

**Figure 1**. Results from network permutation tests, using four different observational and experimental datasets, show significant associations up to between 2 and 4 degrees of separation for a variety of 14 different behaviors and affective states. The Y axis represents the percentage increase in probability that an ego has the trait of interest given that an alter has it, compared with the probability that an ego has the trait given that the alter does not have it. Vertical black lines indicate 95% confidence intervals. For more details, see the related manuscripts cited in the text. Colors indicate data source: yellow: Framingham Heart Study Social Network [14]; blue: AddHealth [1]; green: lab experiment [6]; red: Facebook strong ties [2]; orange: Hadza hunter gatherers [5].

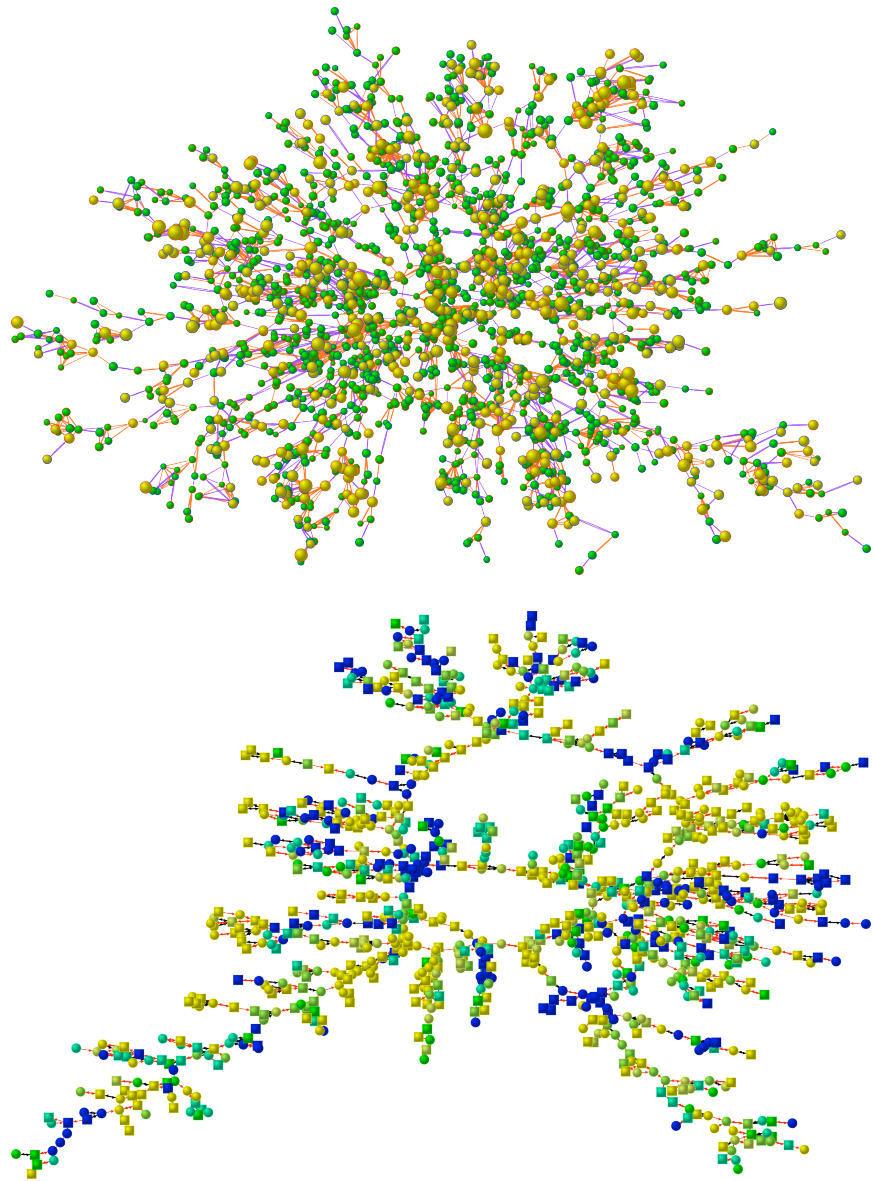

**Figure 2**. Network visualizations showing clustering in obesity (top) and happiness (bottom) in the Framingham Heart Study Social Network in 2000. The top graph shows the largest connected component of friends, spouses, and siblings for whom information about body mass was available. Node border indicates gender (red=female subject, blue=male subject), node color indicates obesity (yellow=BMI>30), node size is proportional to BMI, and tie colors indicate relationship (purple=friend or spouse, orange=family). The bottom graph shows a portion of the largest component of friends, spouses, and siblings for whom information about happiness was available. Each node represents a subject and its shape denotes gender (circles are female, squares are male). Lines between nodes indicate relationship (black for siblings, red for friends and spouses). Node color denotes the mean happiness of the ego and all directly connected (distance 1) alters, with blue shades indicating the least happy, and yellow shades indicating the most happy (shades of green are intermediate). The bottom image involves both 'geodesic smoothing' and sampling, as noted in the text.

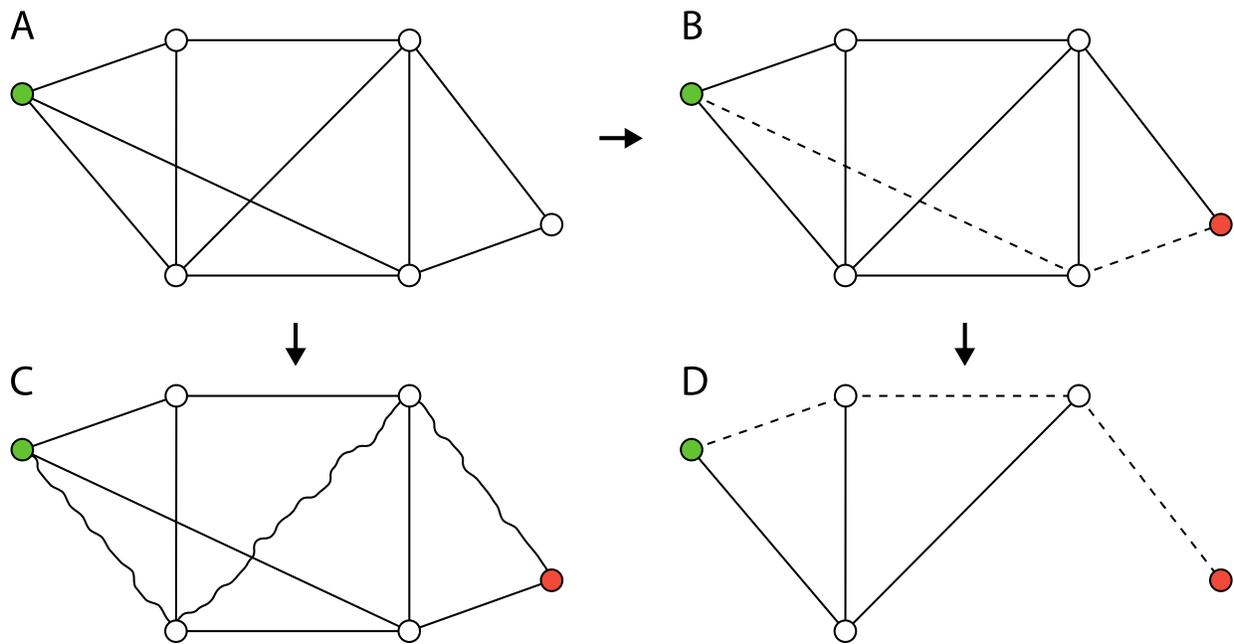

**Figure 3.** Schematic of a network infection and sampling process. (A) The full (unobserved) network with the initially infected node colored green (upper left corner of the network). (B) The shortest path from the source node to the target node colored red (lower right corner of the network) corresponds to the most likely infection path in the fully observed network and has a length of 2. (C) The (unobservable) spreading process unfolds in the (unobserved) network. The actual path taken by the infection is shown with wavy edges. The target node is reached in three steps giving a length of 3. (D) The partially observed network has some nodes and links missing depending on the sampling. The shortest path from source to target has a length of 3 (shown in the dotted lines), corresponding to the length of the most likely path taken by the infection. In this case, using the shortest path length in the fully observed network to estimate the actual path length would result in an underestimate of path length, whereas using the path in the partially observed network, in this case, correctly yields a path length of 3.

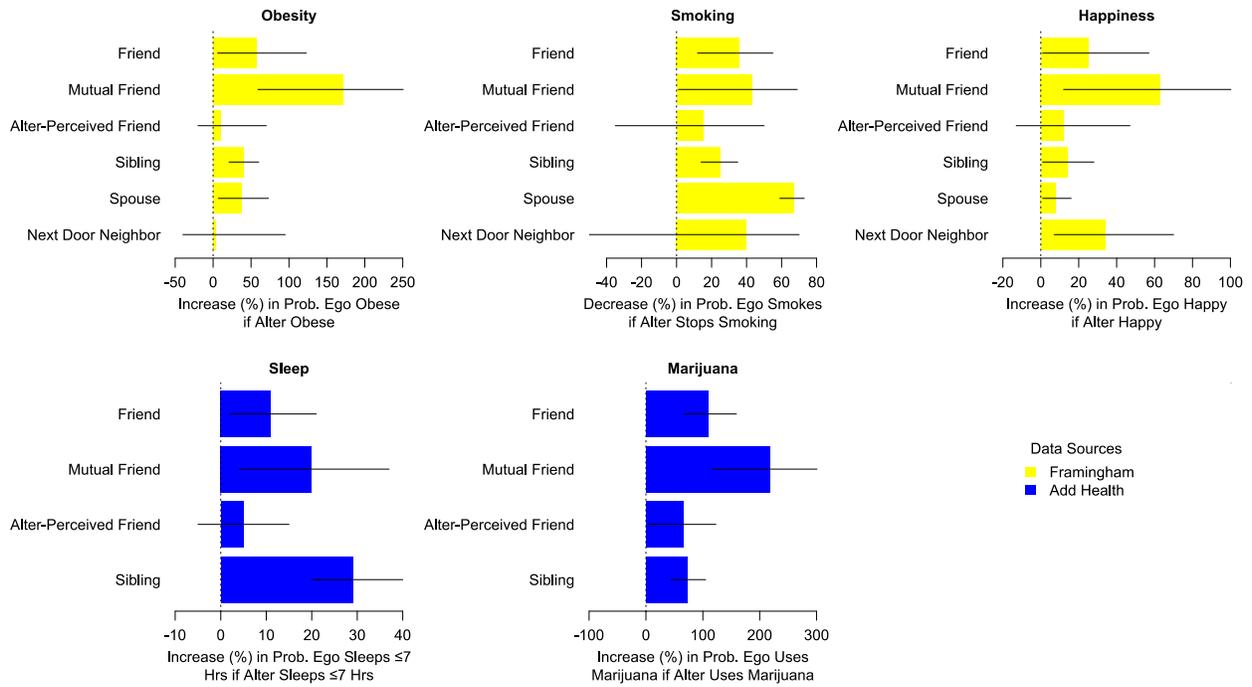

**Figure 4.** Illustrative results from longitudinal regression models for various relationship types and outcomes. Horizontal bars show 95% confidence intervals derived from GEE models by simulating the first difference in alter contemporaneous outcome (changing from 0 to 1) using 1,000 randomly drawn sets of estimates from the coefficient covariance matrix and assuming all other variables were held at their means.

References

1.      Harris KM, Bearman PS, Udry JR. The national longitudinal study of adolescent health: research design. 2010. Available from: http://www.cpc.unc.edu/projects/addhealth/design.
2.      Lewis K, Kaufman J, Gonzalez M, Wimmer A, Christakis N. Tastes, ties, and time: A new social network dataset using Facebook.com. *Social Networks* 2008; **30**: 330-342.
3.      Bond RM, Fariss CJ, Jones JJ, Settle AKE, Marlow C, Fowler JH. A massive scale experiment in social influence and political mobilization. 2012: Under Review.
4.      Christakis NA, Fowler JH. Social network sensors for early detection of contagious outbreaks. *PloS one* 2010; **5**: e12948.
5.      Apicella CL, Marlowe FW, Fowler JH, Christakis NA. Social networks and cooperation in hunter-gatherers. *Nature* 2012; **481**: 497-501.
6.      Fowler JH, Christakis NA. Cooperative behavior cascades in human social networks. *Proceedings of the National Academy of Sciences* 2010; **107**: 5334-5338.
7.      Rand DG, Arbesman S, Christakis NA. Dynamic social networks promote cooperation in experiments with humans. *Proceedings of the National Academy of Sciences* 2011; **108**: 19193-19198.
8.      Fowler JH, Dawes CT, Christakis NA. Model of genetic variation in human social networks. *Proceedings of the National Academy of Sciences* 2009; **106**: 1720-1724.
9.      O'Malley AJ, Christakis NA. Longitudinal analysis of large social networks: Estimating the effect of health traits on changes in friendship ties. *Statistics in Medicine* 2011; **30**: 950-964.
10.     Christakis NA, Fowler JH, Imbens GW, Kalyanaraman K. An empirical model for strategic network formation. *National Bureau of Economic Research Working Paper Series* 2010: 16039.
11.     Onnela J-P, Christakis NA. Spreading paths in partially observed social networks. *Physical Review E* 2012: In Press.
12.     Onnela J-P, Arbesman S, González MC, Barabási A-L, Christakis NA. Geographic constraints on social network groups. *PloS one* 2011; **6**: e16939.
13.     Fowler JH, Settle JE, Christakis NA. Correlated genotypes in friendship networks. *Proceedings of the National Academy of Sciences* 2011; **108**: 1993-1997.
14.     Christakis NA, Fowler JH. The spread of obesity in a large social network over 32 years. *New England Journal of Medicine* 2007; **357**: 370-379.
15.     Fowler JH, Christakis NA. Estimating peer effects on health in social networks: A response to Cohen-Cole and Fletcher; and Trogdon, Nonnemaker, and Pais. *Journal of Health Economics* 2008; **27**: 1400-1405.
16.     Christakis NA, Fowler JH. The collective dynamics of smoking in a large social network. *New England Journal of Medicine* 2008; **358**: 2249-2258.
17.     Rosenquist JN, Murabito J, Fowler JH, Christakis NA. The spread of alcohol consumption behavior in a large social network. *Annals of Internal Medicine* 2010; **152**: 426-433.
18.     Keating NL, O'Malley AJ, Murabito JM, Smith KP, Christakis NA. Minimal social network effects evident in cancer screening behavior. *Cancer* 2011; **117**: 3045-3052.

19. Fowler JH, Christakis N. Dynamic spread of happiness in a large social network: longitudinal analysis over 20 years in the Framingham heart study. *British Medical Journal* 2008; **337**: a2338.

20. Cacioppo JT, Fowler JH, Christakis NA. Alone in the crowd: the structure and spread of loneliness in a large social network. *Journal of Personality and Social Psychology* 2009; **97**: 977-991.

21. Rosenquist JN, Fowler JH, Christakis NA. Social network determinants of depression. *Molecular Psychiatry* 2011; **16**: 273-281.

22. Mednick SC, Christakis NA, Fowler JH. The spread of sleep loss influences drug use in adolescent social networks. *PloS one* 2010; **5**: e9775.

23. McDermott R, Fowler JH, Christakis NA. Breaking up is hard to do, unless everyone else is doing it too: social network effects on divorce in a longitudinal sample followed for 32 years. *SSRN eLibrary* 2009. Available from: http://ssrn.com/paper=1490708.

24. Pachucki MA, Jacques PF, Christakis NA. Social network concordance in food choice among spouses, friends, and siblings. *American Journal of Public Health* 2011; **101**: 2170-2177.

25. Brakefield TA, Mednick SC, Wilson HW, deNeve JE, Christakis NA, Fowler JH. Sexual orientation does not spread in adolescent social networks. 2012: Submitted.

26. Fowler JH, Christakis NA. On Facebook, a picture is worth a thousand words. 2009. Available from: http://www.connectedthebook.com/pages/links/tastes_on_facebook.html.

27. Christakis NA, Fowler JH. *Connected: The Surprising Power of our Social Networks and How They Shape Our Lives.* Little, Brown and Co.: New York, 2009.

28. Smith KP, Christakis NA. Social networks and health. *Annual Review of Sociology* 2008; **34**: 405-429.

29. Wasserman S, Faust K. *Social Network Analysis: Methods and Applications.* Cambridge University Press: Cambridge, U.K., 1994.

30. Jackson MO. *Social and Economic Networks.* Princeton University Press: Princeton, N.J., 2008.

31. Goyal S. *Connections: An Introduction to the Economics of Networks.* Princeton University Press: Princeton, N.J., 2007.

32. O'Malley A, Marsden P. The analysis of social networks. *Health Services and Outcomes Research Methodology* 2008; **8**: 222-269.

33. Newman MEJ. The structure and function of complex networks. *SIAM Review* 2003; **45**: 167-256.

34. Newman MEJ. *Networks: An Introduction.* Oxford University Press: Oxford, U.K., 2010.

35. Easley D, Kleinberg J. *Networks, Crowds, and Markets: Reasoning about a Highly Connected World.* Cambridge University Press: Cambridge, U.K., 2010.

36. Kolaczyk ED. *Statistical Analysis of Network Data: Methods and Models.* Springer: New York, N.Y., 2009.

37. Dawber TR. *The Framingham Study: The Epidemiology of Atherosclerotic Disease.* Harvard University Press: Cambridge, Mass., 1980.

38. Feinleib M, Kannel WB, Garrison RJ, McNamara PM, Castelli WP. The framingham offspring study. Design and preliminary data. *Preventive Medicine* 1975; **4**: 518-525.

39. Campbell KE, Lee BA. Name generators in surveys of personal networks. *Social Networks* 1991; **13**: 203-221.

40.     Marsden PV. Recent developments in network measurement. In *Models and Methods in Social Network Analysis*, Carrington PJ, Scott J, Wasserman S (eds). Cambridge University Press: Cambridge, U.K., 2005.

41.     Marsden PV. Core discussion networks of Americans. *American Sociological Review* 1987; **52**: 122-131.

42.     McPherson M, Smith-Lovin L, Brashears ME. Social isolation in America: changes in core discussion networks over two decades. *American Sociological Review* 2006; **71**: 353-375.

43.     O'Malley AJ, Arbesman S, Steiger DM, Fowler JH, Christakis NA. Egocentric social network structure, health, and pro-social behaviors in a national panel study of Americans. 2012: Under Review.

44.     Szábo G, Barabási A-L. Network effects in service usage. 2007. Available from: http://lanl.arxiv.org/abs/physics/0611177.

45.     Hubert LJ. *Assignment Methods in Combinatorial Data Analysis.* M. Dekker: New York, 1987.

46.     Dekker D, Krackhardt D, Snijders T. Sensitivity of MRQAP Tests to Collinearity and Autocorrelation Conditions. *Psychometrika* 2007; **72**: 563-581.

47.     Thomas AC. Censoring out-degree compromises inferences of social network contagion and autocorrelation. 2010. Available from: http://arxiv.org/abs/1008.1636.

48.     Fisher RA. On the interpretation of X2 from contingency tables, and the calculation of P. *Journal of the Royal Statistical Society* 1922; **85**: 87-94.

49.     Lyons R. The spread of evidence-poor medicine via flawed social-network analysis. 2010. Available from: http://arxiv.org/abs/1007.2876.

50.     McPherson M, Smith-Lovin L, Cook JM. Birds of a feather: homophily in social networks. *Annual Review of Sociology* 2001; **27**: 415-444.

51.     Valente TW. Models and methods for innovation diffusion. In *Models and Methods in Social Network Analysis*, Carrington PJ, Scott J, Wasserman S (eds). Cambridge University Press: Cambridge, U.K., 2005.

52.     Aral S, Muchnik L, Sundararajan A. Distinguishing influence-based contagion from homophily-driven diffusion in dynamic networks. *Proceedings of the National Academy of Sciences* 2009; **106**: 21544-21549.

53.     Gorin AA, Wing RR, Fava JL, Jakicic JM, Jeffery R, West DS, Brelje K, DiLillo VG. Weight loss treatment influences untreated spouses and the home environment: evidence of a ripple effect. *International Journal of Obesity* 2008; **32**: 1678-1684.

54.     Nickerson DW. Is voting contagious? evidence from two field experiments. *American Political Science Review* 2008; **102**: 49-57.

55.     Moody J, McFarland D, Bender-deMoll S. Dynamic network visualization. *American Journal of Sociology* 2005; **110**: 1206-1241.

56.     Hastie T, Tibshirani R, Friedman JH. *The Elements of Sstatistical Learning: Data Mining, Inference, and Prediction.* Springer: New York, NY, 2009.

57.     Kamada T, Kawai S. An algorithm for drawing general undirected graphs. *Information Processing Letters* 1989; **31**: 7-15.

58.     Fruchterman TMJ, Reingold EM. Graph drawing by force-directed placement. *Software: Practice and Experience* 1991; **21**: 1129-1164.

59.     Travers J, Milgram S. An experimental study in the small world problem. *Sociometry* 1969; **35**: 425-443.

60. Dodds PS, Muhamad R, Watts DJ. An experimental study of search in global social networks. *Science* 2003; **301**: 827-829.

61. Brown JJ, Reingen PH. Social ties and word of mouth referral behavior. *Journal of Consumer Research* 1987; **14**: 350-362.

62. Singh J. Collaborative networks as determinants of knowledge diffusion patterns. *Management Science* 2005; **51**: 756-770.

63. Bliss CA, Kloumann IM, Harris KD, Danforth CM, Dodds PS. Twitter reciprocal reply networks exhibit assortativity with respect to happiness. 2011. Available from: http://arxiv.org/abs/1112.1010.

64. Liang K-Y, Zeger SL. Longitudinal data analysis using generalized linear models. *Biometrika* 1986; **73**: 13-22.

65. Elwert F, Christakis N. Wives and ex-wives: A new test for homogamy bias in the widowhood effect. *Demography* 2008; **45**: 851-873.

66. Schildcrout JS, Heagerty PJ. Regression analysis of longitudinal binary data with time-dependent environmental covariates: bias and efficiency. *Biostatistics* 2005; **6**: 633-652.

67. Beck N. Time-series–cross-section data: what have we learned in the past few years? *Annual Review of Political Science* 2001; **4**: 271-293.

68. Banerjee A, Dolado JJ, Galbraith JW, Hendry D. Co-integration, error correction, and the econometric analysis of non-stationary data. *OUP Catalogue* 1993.

69. De Boef S, Keele L. Taking time seriously. *American Journal of Political Science* 2008; **52**: 184-200.

70. VanderWeele TJ, Ogburn EL, Tchetgen EJ. Why and when "flawed" social network analyses still yield valid tests of no contagion. *Statistics, Politics, and Policy* 2012; **3**: 1-11.

71. Steglich C, Snijders TAB, West P. Applying SIENA: an illustrative analysis of the eo-evolution of adolescents' friendship networks, taste in music, and alcohol consumption. *Methodology: European Journal of Research Methods for the Behavioral and Social Sciences* 2006; **2**: 48-56.

72. Steglich C, Snijders TAB, Pearson M. Dynamic networks and behavior: separating selection from influence. *Sociological Methodology* 2010; **40**: 329-393.

73. Mercken L, Snijders TAB, Steglich C, Vertiainen E, De Vries H. Smoking-based selection and influence in gender-segregated friendship networks: a social network analysis of adolescent smoking. *Addiction* 2010; **105**: 1280-1289.

74. Mercken L, Snijders TAB, Steglich C, Vartiainen E, de Vries H. Dynamics of adolescent friendship networks and smoking behavior. *Social Networks* 2010; **32**: 72-81.

75. Lewis K, Gonzalez M, Kaufman J. Social selection and peer influence in an online social network. *Proceedings of the National Academy of Sciences* 2012; **109**: 68-72.

76. Manski CF. Identification of endogenous social effects: the reflection problem. *The Review of Economic Studies* 1993; **60**: 531-542.

77. Shalizi CR, Thomas AC. Homophily and contagion are generically confounded in observational social network studies. *Sociological Methods & Research* 2011; **40**: 211-239.

78. Noel H, Nyhan B. The "unfriending" problem: The consequences of homophily in friendship retention for causal estimates of social influence. *Social Networks* 2011; **33**: 211-218.

79. Fowler JH, Heaney MT, Nickerson DW, Padgett JF, Sinclair B. Causality in political networks. *American Politics Research* 2011; **39**: 437-480.

80. Smith GCS, Pell JP. Parachute use to prevent death and major trauma related to gravitational challenge: systematic review of randomised controlled trials. *BMJ* 2003; **327**: 1459-1461.

81. Cohen-Cole E, Fletcher JM. Detecting implausible social network effects in acne, height, and headaches: longitudinal analysis. *BMJ* 2008; **337**.

82. VanderWeele TJ. Sensitivity analysis for contagion effects in social networks. *Sociological Methods & Research* 2011; **40**: 240-255.

83. Strully KW, Fowler JH, Murabito JM, Benjamin EJ, Levy D, Christakis NA. Aspirin use and cardiovascular events in social networks. *Social Science & Medicine* 2012; **74**: 1125-1129.

84. Visscher PM, Medland SE, Ferreira MAR, Morley KI, Zhu G, Cornes BK, Montgomery GW, Martin NG. Assumption-free estimation of heritability from genome-wide identity-by-descent sharing between full siblings. *PLoS Genetics* 2006; **2**: e41.

85. VanderWeele TJ, Arah OA. Bias formulas for sensitivity analysis of unmeasured confounding for general outcomes, treatments, and confounders. *Epidemiology* 2011; **22**: 42-52.

86. Mercken L, Candel M, Willems P, De Vries H. Disentangling social selection and social influence effects on adolescent smoking: the importance of reciprocity in friendships. *Addiction* 2007; **102**: 1483-1492.

87. Anagnostopoulos A, Kumar R, Mahdian M. Influence and correlation in social networks. *Proceedings of the 14th ACM SIGKDD International Conference on Knowledge Discovery and Data Mining* 2008: 7-15.

88. Bramoullé Y, Djebbari H, Fortin B. Identification of peer effects through social networks. *Journal of Econometrics* 2009; **150**: 41-55.

89. Iwashyna T, Gebremariam A, Hutchins M, Lee J. Can longitudinal GEE models distinguish network influence and homophily? An agent-based modeling approach to measurement characteristics. 2012: Under Review.

90. Block JP, Christakis NA, O'Malley AJ, Subramanian SV. Proximity to food establishments and body mass index in the framingham heart study offspring cohort over 30 years. *American Journal of Epidemiology* 2011; **174**: 1108-1114.

91. Hidalgo CA, Blumm N, Barabási A-L, Christakis NA. A dynamic network approach for the study of human phenotypes. *PLoS Computer Biology* 2009; **5**: e1000353.

92. Fowler JH. Connecting the congress: a study of cosponsorship networks. *Political Analysis* 2006; **14**: 456-487.

93. Fowler JH. Legislative cosponsorship networks in the U.S. House and Senate. *Social Networks* 2006; **28**: 454-465.

94. Fowler JH, Jeon S. The authority of Supreme Court precedent. *Social Networks* 2008; **30**: 16-30.

95. Fowler JH, Johnson TR, Spriggs JF, Jeon S, Wahlbeck PJ. Network analysis and the law: measuring the legal importance of precedents at the U.S. Supreme Court. *Political Analysis* 2007; **15**: 324-346.

96. Barnes MG, Smith TG, Yoder J. Economic insecurity and the spread of obesity in social networks. *Social Science Research Network* 2010. Available from: http://papers.ssrn.com/sol3/papers.cfm?abstract_id=1552185.

97.     Brown H, Hole AR, Roberts J. Going the same "weigh": spousal correlations in obesity in the U.K. *Department of Economics, University of Sheffield, U.K.* 2010: Working Paper.

98.     de la Haye K, Robins G, Mohr P, Wilson C. Obesity-related behaviors in adolescent friendship networks. *Social Networks* 2010; **32**: 161-167.

99.     Trogdon JG, Nonnemaker J, Pais J. Peer effects in adolescent overweight. *Journal of Health Economics* 2008; **27**: 1388-1399.

100.    Halliday TJ, Kwak S. Weight gain in adolescents and their peers. *Economics & Human Biology* 2009; **7**: 181-190.

101.    Carrell SE, Hoekstra M, West JE. Is poor fitness contagious?: Evidence from randomly assigned friends. *Journal of Public Economics* 2011; **95**: 657-663.

102.    McFerran B, Dahl DW, Fitzsimons GJ, Morales AC. Might an overweight waitress make you eat more? How the body type of others is sufficient to alter our food consumption. *Journal of Consumer Psychology* 2010; **20**: 146-151.

103.    Madan A, Moturu ST, Lazer D, Pentland A. Social sensing: obesity, unhealthy eating and exercise in face-to-face networks. *Wireless Health 2010* 2010: 104-110.

104.    McCaffery JM, Franz CE, Jacobson K, Leahey TM, Xian H, Wing RR, Lyons MJ, Kremen WS. Effects of social contact and zygosity on 21-y weight change in male twins. *The American Journal of Clinical Nutrition* 2011; **94**: 404-409.

105.    Leahey TM, LaRose JG, Fava JL, Wing RR. Social influences are associated with BMI and weight loss intentions in young adults. *Obesity* 2011; **19**: 1157-1162.

106.    Leahey TM, Crane MM, Pinto AM, Weinberg B, Kumar R, Wing RR. Effect of teammates on changes in physical activity in a statewide campaign. *Preventive Medicine* 2010; **51**: 45-49.

107.    Hruschka DJ, Brewis AA, Wutich A, Morin B. Shared norms and their explanation for the social clustering of obesity. *American Journal of Public Health* 2011; **101**: S295.

108.    Centola D. The spread of behavior in an online social network experiment. *Science* 2010; **329**: 1194-1197.

109.    Centola D. An experimental study of homophily in the adoption of health behavior. *Science* 2011; **334**: 1269-1272.

110.    de la Haye K, Robins G, Mohr P, Wilson C. Homophily and contagion as explanations for weight similarities among adolescent friends. *Journal of Adolescent Health* 2011; **49**: 421-427.

111.    Campbell MC, Mohr GS. Seeing is eating: how and when activation of a negative stereotype increases stereotype-conducive behavior. *Journal of Consumer Research* 2011; **38**: 431-444.

112.    Burke MA, Heiland FW, Nadler CM. From "overweight" to "about right": evidence of a generational shift in body weight norms. *Obesity* 2010; **18**: 1226-1234.

113.    Hatfield E, Cacioppo JT, Rapson RL. *Emotional Contagion.* Cambridge University Press: Cambridge, U.K., 1994.

114.    Knight J, Gunatilaka R. The rural-urban divide in China: income but not happiness? *Journal of Development Studies* 2010; **46**: 506-534.

115.    Eisenberg D, Golberstein E, Whitelock JL, Downs MF. Social contagion of mental health: evidence from college roomates. 2010: Working Paper.

116.    Schwarze J, Winkelmann R. Happiness and altruism within the extended family. *Journal of Population Economics* 2011; **24**: 1033-1051.

117.    Ali MM, Dwyer DS. Estimating peer effects in adolescent smoking behavior: a longitudinal analysis. *Journal of Adolescent Health* 2009; **45**: 402-408.

118.    Cobb NK, Graham AL, Abrams DB. Social network structure of a large online community for smoking cessation. *American Journal of Public Health* 2010; **100**: 1282-1289.

119.    Mermelstein RJ, Colvin PJ, Klingemann SD. Dating and changes in adolescent cigarette smoking: Does partner smoking behavior matter? *Nicotine & Tobacco Research* 2009; **11**: 1226-1230.

120.    Valente TW, Hoffman BR, Ritt-Olson A, Lichtman K, Johnson CA. Effects of a social-network method for group assignment strategies on peer-led tobacco prevention programs in schools. *American Journal of Public Health* 2003; **93**: 1837-1843.

121.    Valente TW, Ritt-Olson A, Stacy A, Unger JB, Okamoto J, Sussman S. Peer acceleration: effects of a social network tailored substance abuse prevention program among high-risk adolescents. *Addiction* 2007; **102**: 1804-1815.

122.    Campbell R, Starkey F, Holliday J, Audrey S, Bloor M, Parry-Langdon N, Hughes R, Moore L. An informal school-based peer-led intervention for smoking prevention in adolescence (ASSIST): a cluster randomised trial. *The Lancet* 2008; **371**: 1595-1602.

123.    Horton JJ, Rand DG, Zeckhauser RJ. The online laboratory: conducting experiments in a real labor market. *National Bureau of Economic Research Working Paper Series* 2010: 15961.

124.    Aral S, Walker D. Creating Social Contagion Through Viral Product Design: A Randomized Trial of Peer Influence in Networks. *Management Science* 2011; **57**: 1623-1639.

125.    O'Malley AJ, Rosenquist JN, Zaslavsky AM, Christakis NA. Estimation of peer effects using longitudinal data and genetic alleles as instrumental variables. 2012: Working Paper.

126.    Lazer D, Pentland A, Adamic L, Aral S, Barabási A-L, Brewer D, Christakis N, Contractor N, Fowler J, Gutmann M, Jebara T, King G, Macy M, Roy D, Van Alstyne M. Computational Social Science. *Science* 2009; **323**: 721-723.